# Giant spin transfer torque in atomically thin magnetic bilayers

Weihao Cao[1,2], Matisse Wei-Yuan Tu[1,3#], Jiang Xiao[4,5], Wang Yao[1,3*]

[1]Department of Physics, University of Hong Kong, Hong Kong, China

[2]Department of Physics, University of California San Diego, La Jolla, CA 92093-0319, USA

[3]HKU-UCAS Joint Institute of Theoretical and Computational Physics at Hong Kong, China

[4]Department of Physics and State Key Laboratory of Surface Physics, Fudan University, Shanghai 200433, China

[5]Institute for Nanoelectronics Devices and Quantum Computing, Fudan University, Shanghai 200433, China

*Correspondence to: wangyao@hku.hk

# Correspondence to: kerustemiro@gmail.com

**Abstract**

In cavity quantum electrodynamics, the multiple reflections of a photon between two mirrors defining a cavity is exploited to enhance the light-coupling of an intra-cavity atom. We show that this paradigm for enhancing the interaction of a flying particle with a localized object can be generalized to spintronics based on van der Waals 2D magnets. Upon tunneling through a magnetic bilayer, we find the spin transfer torques per electron incidence can become orders of magnitude larger than $\hbar/2$, made possible by electron's multi-reflection path through the ferromagnetic monolayers as an intermediate of their angular momentum transfer. Over a broad energy range around the tunneling resonances, the damping-like spin transfer torque per electron tunneling features a universal value of $\frac{\hbar}{2}\tan\frac{\theta}{2}$, depending only on the angle $\theta$ between the magnetizations. These findings expand the scope of magnetization manipulations for high-performance and high-density storage based on van der Waals magnets.

**Introduction**

Magnetic information storage technology exploits the magnetization direction of ferromagnet to encode information, an example of which is the magnetic random-access-memory (MRAM) based on magnetic tunnel junctions (MTJ)[1,2]. An MTJ consists of two ferromagnetic layers: a fixed reference layer and a free switching layer. The transfer of spin angular momentum between the ferromagnets and the passing electrons realizes both reading and writing functions in an MTJ. Through such transfer, a fixed ferromagnet acting as a polarizing layer aligns the spin of incident electrons, which, in passing through a free magnetic layer, can have two distinct values of resistance depending on the relative alignment of the magnetizations[1]. This allows electric reading of the magnetic information. In turn, the angular momentum of the electron can be transferred to the magnet as a back-action (Fig. 1a), a phenomenon known as spin-transfer torque [2-5], which is actively explored for all electric manipulation of magnetizations (writing) in storage devices[2,6].

The recent discovery of atomically thin 2D magnets[7-14] has provided an ideal platform for spintronic applications. A great asset of these van der Waals magnets is the ability to engineer magnetic multilayer devices with ultimate miniaturization in thickness, implying the potential to further shrink device sizes for high density storage. And the atomically sharp and clean interface[15] between magnetic and non-magnetic sublayers is advantageous for removing device fluctuations and imperfections. Preliminary forms of atomically thin van der Waals spintronic devices are first demonstrated using chromium trihalides[16-20]. The three-atom thick monolayer is a ferromagnet, a thinnest possible unit to encode magnetic information, while adjacent layers separated by van der Waals gaps can be arranged in parallel or anti-parallel configurations through the antiferromagnet interlayer coupling and magnetic field. Sandwiched between graphene contacts, the few-layer chromium trihalides function as multiple-spin-filter MTJ, and record-high tunneling magnetoresistance ratio are measured[16-20], highly favored for the reading. Manipulation of magnetization orientations in these few-layer magnets by electrostatic gating has also been demonstrated[21-23]. Concerning the writing controls, van der Waals magnet has another significance: the transmission through the magnetic multilayer of atomic thickness can be fully quantum coherent, which is a new realm to explore the current induced torques.

In conventional MTJ devices[24-27], the angular momentum transfer per electron incidence from the polarizing layer to the free layer is upper bounded by $\hbar/2$, the maximum amount an electron can carry (Fig. 1a). Here we show that the quantum coherent tunneling through atomically thin magnetic bilayers can lead to an analog of optical microcavity for enhancing spin transfer torque (STT), well exceeding the above limit (Fig. 1b). The scattering wavefunctions at the tunneling resonances have a spin current density profile peaked between the two ferromagnetic monolayers, resembling the intra-cavity field for a resonant photon. As a result, the torque per electron incidence becomes orders of magnitude larger than $\hbar/2$. Electron's role as an intermediate for angular momentum transfer is radically amplified through its multiple reflections between the two monolayers (Fig. 1b). We find a field-like torque that is equivalent to the RKKY interaction mediated by the quasi-bound electron, as well as a damping-like STT that tends to align downstream layer parallel to upstream layer, and upstream layer anti-parallel to downstream one. When we turn to the damping-like torque per electron transmission, a figure of merit for the current driven torque, a universal value of $\frac{\hbar}{2}\tan\frac{\theta}{2}$ is found over a broad range of incident energy, depending only on the angle $\theta$ between the two magnetizations. This finding points to highly efficient manipulation of 2D magnets by tunneling current for spintronics applications.

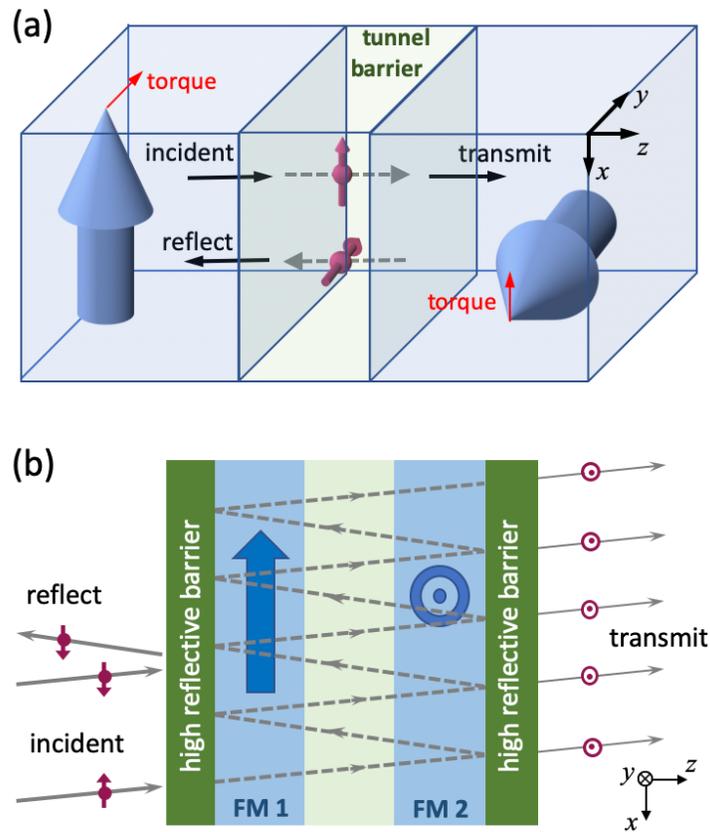

**Figure 1 | Cavity enhancement of spin transfer torque. a,** In a conventional magnetic tunnel junction, the angular momentum transfer per electron incidence from the polarizing layer to the free layer is upper bounded by electron's angular momentum $\hbar/2$. **b,** Schematic of two high reflective barriers sandwiching the atomically thin magnetic bilayer. The multiple reflections can radically amplify electron's role as an intermediate of angular momentum transfer, and the torque on each magnetic layer can well exceed $\hbar/2$ per electron incidence.

**Method of analysis**

Fig. 2a schematically shows an exemplary van der Waals (vdW) tunnel junction device consisting of a magnetic bilayer sandwiched between graphene/graphite contacts. The plane of a magnetic monolayer is defined to be the *x-y* plane (Fig. 2b) so the vertical tunneling direction is the *z*-direction. Without losing generality, we set the magnetization of layer 1, $M_1$, in the -*z* direction, and the magnetization of layer 2, $M_2$, in *x-z* plane with angle $\theta$ from $M_1$, c.f. Fig. 2b. In a ferromagnetic monolayer, the majority and the minority spin states are separated by a large energy gap of hundreds of meV[16]. The majority (minority) spin in layer 1 is denoted by $|\downarrow\rangle$ ($|\uparrow\rangle$). Tilted by an angle $\theta$, the majority spin in layer 2 is thus $|\theta\rangle = \cos\left(\frac{\theta}{2}\right)|\downarrow\rangle - \sin\left(\frac{\theta}{2}\right)|\uparrow\rangle$. We first concentrate on the tunneling transport facilitated by the majority spin states in the two wells, while the effects of minority spin states will be included later. Such tunneling transport can be described by the Hamiltonian: $\hat{H} = \sum_k \left(E_1(\boldsymbol{k})\hat{a}_{1,\boldsymbol{k}}^\dagger \hat{a}_{1,\boldsymbol{k}} + E_2(\boldsymbol{k})\hat{a}_{2,\boldsymbol{k}}^\dagger \hat{a}_{2,\boldsymbol{k}}\right) + \sum_l E_l\, \hat{a}_{l,\sigma}^\dagger \hat{a}_{l,\sigma} + \sum_r E_r\, \hat{a}_{r,\sigma}^\dagger \hat{a}_{r,\sigma} + \left[\sum_{l,\boldsymbol{k}} g_{l,\boldsymbol{k}}\hat{a}_{l,\downarrow}\hat{a}_{1,\boldsymbol{k}}^\dagger + \sum_{r,\boldsymbol{k}} g_{r,\boldsymbol{k}}\hat{a}_{2,\boldsymbol{k}}\hat{a}_{r,\theta}^\dagger + \sum_{\boldsymbol{k}} t_{12}(\boldsymbol{k})\hat{a}_{1,\boldsymbol{k}}\hat{a}_{2,\boldsymbol{k}}^\dagger + \text{h.c.}\right]$ Here $\hat{a}_{1,\boldsymbol{k}}^\dagger$ and $\hat{a}_{2,\boldsymbol{k}}^\dagger$ create carrier with in-plane momentum $\boldsymbol{k}$ in the first layer with spin $|\downarrow\rangle$ and second layer with spin $|\theta\rangle$ respectively. Since the in-plane momentum is conserved between the two lattice-matched magnetic layers, their interlayer hopping couples only states of the same $\boldsymbol{k}$ from the two layers. Interlayer hopping also conserves spin, so the coupling is a function of the angle between the magnetizations, i.e. $t_{12}(\boldsymbol{k}) = t(\boldsymbol{k})\cos\left(\frac{\theta}{2}\right)$, as the two quantum well states have spin configuration $|\downarrow\rangle$ and $|\theta\rangle$ respectively. For the anti-parallel configuration ($\theta = \pi$), $t_{12}$ vanishes, while at other $\theta$ value, it leads to layer hybridization of the spin majority states. $\hat{a}_{l,\sigma}^\dagger$ ($\hat{a}_{r,\sigma}^\dagger$) create the continuum modes in the left (right) contact of spinor $\sigma$. Since the tunneling between the magnetic layer and the adjacent contact is spin conserving, the state in the first (second) magnet couples to the left (right) contact states of the same spin configuration only, denoted as $\hat{a}^\dagger_{l,\downarrow}$ ($\hat{a}^\dagger_{r,\theta}$). Without lattice matching between the contact materials and the magnetic layers, the quantum well state of a given $\boldsymbol{k}$ in the magnetic layer can be coupled to a continuum of modes in the contact.

At every $\boldsymbol{k}$, we thus have an effective one-dimensional tunneling channel, characterized by the 1D Hamiltonian $\widehat{H}_{\boldsymbol{k}} = E_1(\boldsymbol{k})\hat{a}^\dagger_1 \hat{a}_1 + E_2(\boldsymbol{k})\hat{a}^\dagger_2 \hat{a}_2 + \sum_l E_l\, \hat{a}^\dagger_{l,\sigma}\hat{a}_{l,\sigma} + \sum_r E_r\, \hat{a}^\dagger_{r,\sigma}\hat{a}_{r,\sigma} + \left(\sum_l g_{l,\boldsymbol{k}} \hat{a}_{l,\downarrow}\hat{a}^\dagger_1 + t_{12}\hat{a}_1\hat{a}^\dagger_2 + \sum_r g_{r,\boldsymbol{k}} \hat{a}_2 \hat{a}^\dagger_{r,\theta} + \text{h.c.}\right)$. With the strong vertical confinement of carriers in each monolayer and their weak vdW coupling in the heterostructures, for the tunneling transport, we model this device as coupled quantum wells as illustrated in Fig. 2b. Each monolayer corresponds to a quantum well with spin-dependent potential profile determined by the magnetization direction. The two magnetic quantum wells are separated from each other and from the contacts by barriers that correspond to the vdW gaps. The barrier profile is spin independent, and tunneling between the lattice-matched bilayer conserves in-plane momentum. The latter effectively reduces the tunneling transport to a one-dimensional problem, where the coupled quantum wells feature tunneling resonances at the quasi-bound states. With the lattice mismatch between the graphene electrode and the magnetic layer, the continuity of choices of states in the electrodes is translated into the continuously varied incident energies in the one-dimensional problem.

Such a model has been adopted for description of the giant tunneling magneto-resistance through bilayer and few-layer $CrI_3$ where neighboring monolayers are of anti-parallel (parallel) magnetization at zero (finite) field[16], reproducing well the I-V curves when the contact Fermi energies are set below the tunneling resonances. The calculations of the $CrI_3$/graphene heterostructures therein has used the barrier width $d = 2.76\text{Å}$, quantum well width $w = 5.52\text{Å}$, and the spin-splitting $\Delta = 0.5$ eV in each well. The height of the vdW barriers roughly corresponds to the electron affinity, taken to be 4 eV.

We adopt the same set of parameters here in the numerical calculations, while a parameter independent universal value for the STT near the tunneling resonance will be established. We set the potential energy matrix to be $V_0 + \Delta\, \widehat{\sigma}_z$ in the first well, and $V_0 + \Delta\,(\widehat{\sigma}_z \cos\theta + \widehat{\sigma}_x \sin\theta)$ in the second well. The potential profile generated with these parameters is shown in Fig. 2b for an anti-parallel configuration ($\theta = \pi$). The scattering state wavefunction at this spin-dependent potential, $\Psi(z) = \Psi_\downarrow \otimes |\downarrow\rangle + \Psi_\uparrow \otimes |\uparrow\rangle$, is solved by the wavefunction matching between the well, barrier and contact regions. By examining $\langle \hat{\sigma}_i \rangle \equiv \Psi^* \hat{\sigma}_i \Psi$ using the Schrodinger equation, we find $(\hbar/2) d\langle \hat{\sigma}_i \rangle/dt = -\nabla \cdot Q_i$. This then gives $Q_i(z) \equiv \frac{\hbar^2}{2m} \text{Im}\left(\Psi^* \hat{\sigma}_i \otimes \frac{\partial}{\partial z} \Psi\right)$, the spin current propagating in the out-of-plane ($z$) direction for spin component $i = x, y, z$.

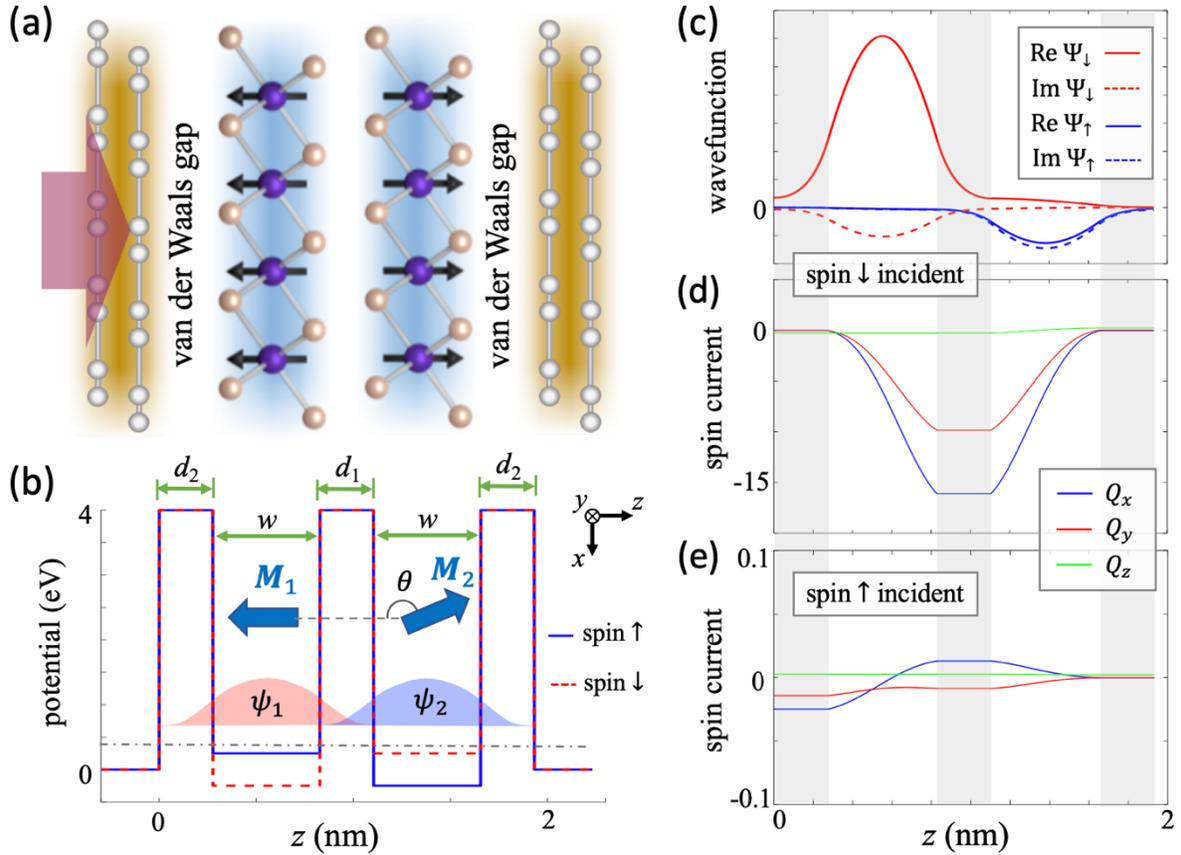

**Figure 2 | Cavity amplification of spin current in coupled double magnetic quantum wells**. **a**, Schematic of a van der Waals heterostructure consisting of two ferromagnetic monolayers, each corresponding to a quantum well, and graphene contacts. **b**, Potential profile of the coupled quantum wells in anti-parallel magnetization configurations ($\theta = \pi$). The barriers correspond to the van der Waals (vdW) gaps. Blue solid (red dashed) line is for spin ↑ (↓) electron. $\psi_1$ and $\psi_2$ denote the wavefunction of quantum well state in the first and second layer respectively. $d_1 = d_2 = 2.76$ Å, and $w = 5.52$ Å. **c**, A scattering wavefunction $\Psi_\downarrow \otimes |\downarrow\rangle + \Psi_\uparrow \otimes |\uparrow\rangle$, for a spin ↓ incidence at tunneling resonance (denoted by dash-dotted line in **b**), under $\theta = 0.99\pi$. The shades mark barrier regions (vdW gaps). **d**, Spin current profiles for the wavefunction in **c**. $Q_i$ denotes the spin current propagating in the $z$ direction carrying spin component $i$, normalized by the incident probability current, and is in unit of $\hbar/2$. **e**, Spin currents for a spin ↑ incidence at the same energy as in **c**.

## Results

In Fig. 2c-e, we show an example of the wavefunction and spin currents calculated under an angle $\theta = 0.99\pi$ between the two magnetizations. Fig. 2c plots the wavefunction at a tunneling resonance for an electron incident on the $|\downarrow\rangle$ state (parallel to the magnetization in layer 1, $M_1$). From the calculations of the transmission probabilities, we see that there is nearly perfect transmission at the concerned energy, while the transmitted electron is found to have its spin parallel to the magnetization of layer 2 ($M_2$). This corroborates the construction of $\widehat{H}_k$ including only the spin majority states in the two layers. The corresponding spin currents are shown in Fig. 2d. $Q_x$ and $Q_y$ both have a standing-wave like profile between the two outer barriers, analogous to the intra-cavity optical field in a microcavity. The maximum spin current is obtained inside the middle barrier between the two monolayer magnets, reaching a value that is 15 times of the incident spin current. In contrast, the spin currents remain small everywhere for a spin ↑ incidence at the same energy (Fig. 2e). From the calculations of the reflection coefficients, we find that the spin ↑ electron is almost fully reflected. For a spin unpolarized incidence, the first magnet can filter out a spin current, which is then resonantly amplified by the multiple reflections through the two magnets.

The spin current profile in Fig. 2d implies sizable torque exerted on each of the magnetic monolayer. The torques on the first and second layer are $\boldsymbol{\tau}^1 = \boldsymbol{Q}(z_1) - \boldsymbol{Q}(z_2)$ and $\boldsymbol{\tau}^2 = \boldsymbol{Q}(z_2) - \boldsymbol{Q}(z_3)$ respectively, where $\boldsymbol{Q} = Q_x\widehat{\boldsymbol{x}} + Q_y\widehat{\boldsymbol{y}} + Q_z\widehat{\boldsymbol{z}}$, and $z_l$ is a coordinate inside the $l$-th barrier. Our calculations find that the torque is always perpendicular to the magnetization of the corresponding layer, $\boldsymbol{\tau}^i \perp \boldsymbol{M}_i$. Each torque can be decomposed into two components perpendicular and parallel to the ($x$-$z$) plane spanned by $\boldsymbol{M}_1$ and $\boldsymbol{M}_2$[6]. The perpendicular components are known as the field-like torques ($\tau_f^{1,2}$), which tend to rotate the magnetizations about each other, keeping $\theta$ unchanged. And the parallel components are known as the damping-like spin transfer torques ($\tau_d^{1,2}$), which tends to change the angle $\theta$. These torque components are schematically illustrated in the inset of Fig. 3.

Figure 3a-c plots the magnitudes of different torque components as functions of angle $\theta$ and incident energy $E_{in}$, for spin-unpolarized incidence, and Fig. 3d shows the corresponding

transmission probability $T$. At a general angle $\theta$, all torque components show two resonant peaks, corresponding respectively to the tunneling resonances at the bonding and anti-bonding states in the coupled double wells (Fig. 3d).

Both the torques and transmission probability vanish for the collinear configuration with $\theta = \pi$. In a general non-colinear configuration, we find that at both tunneling resonances $\tau_d^1$ tends to enlarge the angle $\theta$ between the magnetizations, while $\tau_d^2$ tends to reduce it (Fig. 3a). These qualitative behaviors are consistent with those observed in conventional MTJ and spin valve, where the damping-like STT tends to align downstream layer parallel to upstream layer, and upstream layer anti-parallel to downstream one. For the field-like torque, we find $\boldsymbol{\tau}_f^1 = -\boldsymbol{\tau}_f^2$, so they tend to rotate the plane spanned by $\boldsymbol{M}_1$ and $\boldsymbol{M}_2$ about the axis of $\boldsymbol{M}_1 + \boldsymbol{M}_2$. It is worth noting that the field-like torque has opposite sign at the bonding and the anti-bonding tunneling resonances (Fig. 3b).

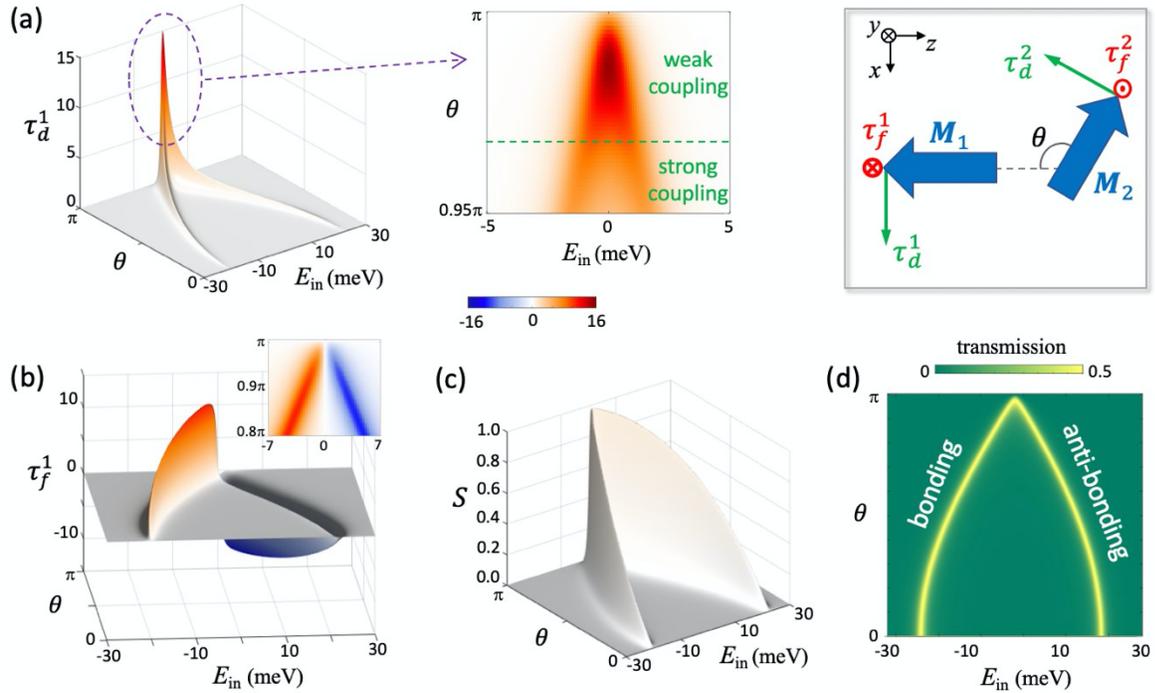

**Figure 3 | Giant spin transfer torque**. The (positive) directions for the torque components are illustrated in the right upper inset. **a**, Magnitude of the damping-like spin transfer torque $\tau_d^1$ on layer 1. $\tau_d^2$ has nearly the same magnitude, and its different direction is noted in the inset. **b**, Field-like torque $\tau_f^1$ on layer 1. The one on layer 2 is exactly its opposite: $\tau_f^2 = -\tau_f^1$. **c**, Angular momentum $S$ carried by each out-going electron, averaged over transmission and reflection. **d**, Transmission probability $T$. All quantities are calculated under a spin-unpolarized incidence. The torques are normalized by the incident probability current, and are given in unit of $\hbar/2$ (i.e. torque value of 1 corresponds to angular momentum transfer of $\hbar/2$ per incident electron). $E_{\text{in}}$ is the incident energy whose reference 0 is set to the middle of the two resonant energies, which corresponds to $(E_1(\mathbf{k}) + E_2(\mathbf{k}))/2$ for the in-plane momentum $\mathbf{k}$ concerned.

The torques plotted in Fig. 3 have been normalized by the incident probability current, so the value can be taken as angular momentum transfer per electron incidence. We find an accelerated increase of the damping-like STT $\tau_d^{1,2}$ as a function of $\theta$, reaching a remarkable peak value of 15 quanta of $\frac{\hbar}{2}$ per electron incidence at $\theta \sim 0.98\pi$, followed by a rapid decay towards $\theta = \pi$. With the different directions of $\boldsymbol{\tau}_d^1$ and $\boldsymbol{\tau}_d^2$ noted ($\perp \boldsymbol{M}_1$ and $\perp \boldsymbol{M}_2$ respectively in the x-z plane), the quantitative behaviors of $\tau_d^1$ and $\tau_d^2$ are nearly the same, and only one is shown in Fig. 3a. In comparison, the field-like torque $\tau_f^{1,2}$ is a smoother function of $\theta$ (Fig. 3b).

To understand the above behaviors in more depth, we inspect the spin current density inside the middle barrier, where the wavefunction is in general a superposition of the evanescent tails of the quantum well states from the two layers (c.f. Fig. 2b), and can be expressed: $\langle z|\Psi\rangle = \beta_1 e^{-\kappa(z-z_0)} \otimes |\downarrow\rangle + \beta_2 e^{\kappa(z-z_0)} \otimes |\theta\rangle$, $z_0$ the coordinate of barrier center. Evaluating the spin currents for this wavefunction, we find

$$Q_x(z_0) = i(\beta_1^*\beta_2 - \beta_1\beta_2^*)\frac{\hbar^2 \kappa}{2m}\sin\left(\frac{\theta}{2}\right), \quad (1)$$

$$Q_y(z_0) = (\beta_1^*\beta_2 + \beta_1\beta_2^*)\frac{\hbar^2 \kappa}{2m}\sin\left(\frac{\theta}{2}\right),$$

$$Q_z(z_0) = -i(\beta_1^*\beta_2 - \beta_1\beta_2^*)\frac{\hbar^2 \kappa}{2m}\cos\left(\frac{\theta}{2}\right).$$

For the isolated coupled wells, $\beta_1$ and $\beta_2$ are real numbers, so $Q_x = Q_z = 0$, and the non-zero $Q_y$ gives rise to a field-like torque which takes opposite sign at the bonding and anti-bonding resonances. Through the Landau-Lifshitz-Gilbert equation[6], one can show that this torque is equivalent to a Hamiltonian term $J\boldsymbol{M}_1 \cdot \boldsymbol{M}_2$, i.e. a RKKY coupling between the two magnetizations mediated by the coupled quantum well state of the electron. The field-like torque we obtained in the scattering state solution under finite coupling to the contacts (Fig. 3b) is a reminiscence of this RKKY coupling. For a scattering state, the associated probability current $j$ corresponds to the transmission probability $T$, and reads,

$$j(z_0) = -i(\beta_1^*\beta_2 - \beta_1\beta_2^*)\frac{\hbar\kappa}{m}\cos\left(\frac{\theta}{2}\right), \qquad (2)$$

when evaluated inside the middle barrier. Its finite value requires the spin currents $Q_z$ and $Q_x$ to be finite as well in the middle barrier. Eq. (1) and (2) together lead to,

$$\frac{\hbar}{2}\tan\frac{\theta}{2} = -\frac{Q_x(z_0)}{j(z_0)} = \frac{\tau_d^1}{T}. \qquad (3)$$

Remarkably, the STT on the upstream magnet per electron transmission, $\tau_d^1/T$, has a universal value independent of the incident energy and the model parameters.

We can define two regimes of resonance tunneling through the coupled wells, depending on the relative strength of their coupling $t_{12}$ with respect to the broadening $\Gamma$ of the tunneling resonance. In the strong coupling regime, i.e. $\Gamma < t_{12} = t\cos\left(\frac{\theta}{2}\right)$, the bonding and anti-bonding resonances are separated as two peaks (Fig. 3d), with an *on-resonance* transmission probability $T_{res} = 0.5$ corresponding to fully transmitted $|\downarrow\rangle$ incidence, and fully blocked $|\uparrow\rangle$. From Eq. (3), $\tau_d^1 = T\frac{\hbar}{2}\tan\frac{\theta}{2}$, i.e. the spin transsfer torque per *incidence* features two resonance peaks accordingly. The on-resonance torque value is then $\frac{\hbar}{4}\tan\frac{\theta}{2}$, in excellent agreement with the rising behavior displayed in Fig. 3a. The small neighborhood of $\theta = \pi$ is the weak coupling regime $t_{12} < \Gamma$, where the two tunneling resonances merge into one (Fig. 3d). In this regime $T_{res}$ drops from 0.5 to 0, which is responsible for the rapid decay of $\tau_d^1$ as $\theta$ approaches $\pi$ (Fig.3a). Fig. 4a plots the on-resonance transmission probability $T_{res}$ and the on-resonance $\tau_d^1$ value as functions of $\theta$, and their ratio (black curve) is found in perfect agreement with the curve (red dashed) dictated by Eq. (3).

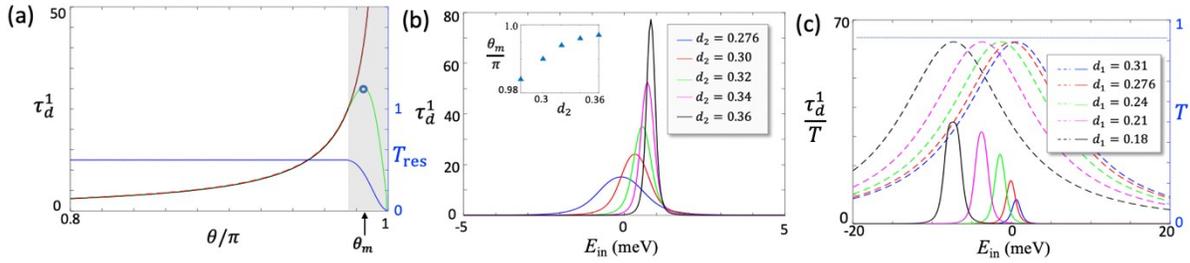

**Figure 4 | Universal value of damping-like spin transfer torque per electron transmission.**
**a**, On resonance spin transfer torque $\tau_d^1$ (green), and transmission probability $T$ (blue). Black curve plots the ratio $\frac{1}{2}\frac{\tau_d^1}{T}$, and red dashed one is the function $\frac{1}{2}\tan\frac{\theta}{2}$ for comparison. $\tau_d^1$ value is given in unit of $\frac{\hbar}{2}$. The shaded area marks the weak coupling regime where $T$ drops from 0.5 (perfect transmission of $\downarrow$ incidence) to 0. **b**, $\tau_d^1$ as function of incident energy $E_{in}$ plotted at the angle $\theta_m$ where the maximum torque value is reached (c.f. **a**). Curves in different color correspond to different outer barrier thickness $d_2$ (c.f. Fig. 2b). The corresponding $\theta_m$ are shown in the inset. **c**, $\frac{\tau_d^1}{T}$ (dashed) and $T$ (solid) as functions of $E_{in}$, when the inner barrier thickness $d_1$ is varied. $\theta = 0.99\pi$ in **c**.

Fig. 4b-c further examine the effects of different quantum-well parameters. In Fig. 4b, under various outer barrier width $d_2$, the damping-like torque $\tau_d^1$ as a function of incident energy is plotted at the angle $\theta_m$ where $\tau_d^1$ reaches its maximum. Wider $d_2$ leads to higher quality factor Q of the tunneling resonance, where an electron has more rounds of reflection between the outer barriers before exiting, hence a larger angular momentum transfer between the two magnetic layers becomes possible. The increase of $\tau_d^1$ peak value with Q is indeed observed in Fig. 4b. We also change the width $d_1$ of the middle barrier, equivalent to the change of the coupling constant $t$ between the two quantum-well states, and the increase of $\tau_d^1$ with $t$ is observed in Fig. 4c. The intuitive trends reflected in Fig. 4b and 4c can also be understood from the perspective offered by Eq. (3). With the increase of either Q or $t$, the weak coupling regime shrinks towards $\theta = \pi$ so that $\tau_d^1$ can pick up higher value on the $\tan\frac{\theta}{2}$ curve.

In Fig. 5a, we allow a detuning $\delta \equiv E_1 - E_2$ between the two wells, which can quench their hybridization and hence the on-resonance transmission. Nevertheless, over a large range of incident energy, the STT per transmission $\frac{\tau_d^1}{T}$ only has modest deviation from the universal value $\frac{\hbar}{2}\tan\frac{\theta}{2}$. This allows a flexible bias window to utilize the sizable STT driven by the tunneling current. In Fig. 5b and 5c, we performed the calculation with asymmetry between the tunneling barriers with the top and bottom contacts, in height and in width respectively. While the transmission probability can be affected by such asymmetry, the spin transfer torque per electron transmission is still well described by the universal value. Indeed, the derivation of Eq. (3) has not relied on any assumption that the tunneling barriers with top and bottom contacts are identical, so the results are applicable in the presence of inevitable asymmetry in realistic devices.

It is interesting to note that the non-colinear magnetization configurations function as a highly efficient spin pump at the tunneling resonances. Fig. 3c plots the angular momentum $\boldsymbol{S}$ carried by each out-going electron, averaged over transmission and reflection. With incidence being unpolarized, $\boldsymbol{S} + \boldsymbol{\tau}_d^1 + \boldsymbol{\tau}_f^1 + \boldsymbol{\tau}_d^2 + \boldsymbol{\tau}_f^2 = 0$, as a result of the angular momentum conservation. Remarkably, $S$ at the tunneling resonances is a sizable fraction of $\hbar/2$ in almost the entire range of $\theta$, meaning that the bilayer has turned an unpolarized incoming flux into a net outgoing spin current. This spin pump is contributed by the transmission part of the electron where the spin configuration is changed from $|\downarrow\rangle$ to $\cos\left(\frac{\theta}{2}\right)|\downarrow\rangle - \sin\left(\frac{\theta}{2}\right)|\uparrow\rangle$, from which we find,

$$\frac{S}{T} = \hbar \cos\frac{\pi - \theta}{2}. \qquad (4)$$

The ratio of $S$ and $T$ from the numerical calculation (Fig. 3c and 3d) well satisfies the above relation. In the strong coupling regime, $T_{\text{res}} = 0.5$, so the on resonance $S = \frac{\hbar}{2}\cos\frac{\pi-\theta}{2}$ (c.f. Fig. 3c), and its upper bound $\frac{\hbar}{2}$ is approximately reached at $\theta \sim 0.97\pi$, the boundary of the strong and weak coupling regimes. In the weak coupling limit, $S \cong \hbar T$, where the on-resonance value drops to 0 at $\theta = \pi$.

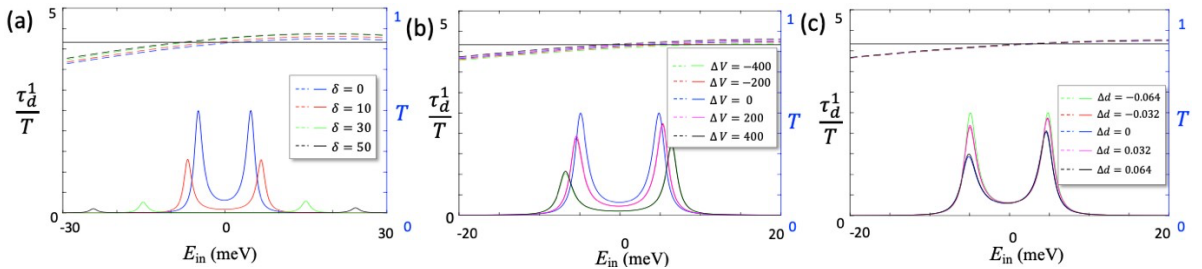

**Figure 5 | Effect of asymmetric configuration on damping-like spin transfer torque per electron transmission.** $\frac{\tau_d^1}{T}$ (dashed) and $T$ (solid) as functions of $E_{\text{in}}$, where curves in different colors correspond to different energy detuning between the two wells $\delta$ in the unit of $meV$ in **a**, different asymmetry between two outer barrier heights in the unit of $meV$ in **b**, and different asymmetry between two outer barrier width in the unit of nm in **c**. In **b**, the two outer barrier heights $V_L$ and $V_R$ are set to $\frac{V_L+V_R}{2} = 4eV$ and $\Delta V = V_R - V_L$. In **c**, the two outer barrier widths $d_L$ and $d_R$ are set to $\frac{d_L+d_R}{2} = 0.276$ nm and $\Delta d = d_R - d_L$. For all plots $\theta = 0.85\pi$, and the the corresponding $\tan\frac{\theta}{2}$ is marked by the constant solid horizontal lines.

**Discussions**

Concerning the 2D nature of the tunneling transport, we restore the in-plane momentum $\bm{k}$ conserved between the two magnetic layers. Referring to $\hat{H}$ introduced in **Method of analysis**, the bonding and anti-bonding resonances are given by $\mathcal{E}_{\pm}(\bm{k}) = \frac{E_1(\bm{k})+E_2(\bm{k})}{2} \pm \sqrt{\left(\frac{E_1(\bm{k})-E_2(\bm{k})}{2}\right)^2 + t_{12}^2}$, and the contributions to the STT and transmission have the same behavior as presented in Fig. 2 to 5, except the overall shift of the resonances. $\mathcal{E}_{\pm}(\bm{k})$ give the dispersions of the lowest energy spin polarized band in the bilayer split by the interlayer coupling $t_{12}$.

The total STT and the tunneling current are given respectively as: $\tau_{\text{sum}} = \int dEd\bm{k}\,[f_l(E) - f_r(E)]\tau_d^1(E,\bm{k})$, and $j = \int dEd\bm{k}\,[f_l(E) - f_r(E)]T(E,\bm{k})$, where $f_{l(r)}(E)$ is the Fermi distribution function in the left (right) contact. When the bias window overlaps with the bands $\mathcal{E}_{\pm}(\bm{k})$, the dominant contributions to $\tau_{\text{sum}}$ and $j$ are the resonant tunneling through a range of $\bm{k}$, where the ratio of on resonance $\tau_d^1(E,\bm{k})$ and $T(E,\bm{k})$ is given by Eq. (3), determined by the angle $\theta$ only. In such case, the damping-like STT per tunneling is still of the universal value:

$$\frac{\tau_{\text{sum}}}{j} = \frac{\hbar}{2}\tan\frac{\theta}{2}. \qquad (5)$$

When the bias window is below the band edge, quantitative drop of the torque from the universal value can be expected with the increase of detuning, as Fig. 4c shows.

These results can be straightforwardly generalized to tunneling through the spin minority states in the quantum well. The only difference lies in the opposite sign of the STT, while the torque magnitude is still described by Eq. (3), if the incident energy is near resonance with the spin minority quantum well state. The overall damping-like STT per transmission is then

$$\frac{\tau_{\text{sum}}}{j} = \frac{j_\downarrow - j_\uparrow}{j_\downarrow + j_\uparrow}\frac{\hbar}{2}\tan\frac{\theta}{2}, \qquad (6)$$

where $j_\downarrow$ ($j_\uparrow$) is the tunneling current for electron incident in $\downarrow$ ($\uparrow$) state, the majority (minority) spin species of the upstream monolayer.

The spin quantization direction can always be chosen according to the magnetization direction, not necessarily the same as the propagation. The spin subspace is decoupled from the coordinate space in the absence of spin-orbit coupling. The results are therefore applicable to arbitrary configurations of the bilayer magnetizations, and to van der Waals magnets of either out-of-plane anisotropy such as CrI$_3$[7], Fe$_3$GeTe$_2$[9,10], or in-plane anisotropy as the lately discovered CrCl$_3$[28-30]. Versatile current induced magnetization dynamics can be expected from the interplay of the giant STT with the distinct magnetization anisotropy and interlayer magnetic coupling in various 2D magnets.

We also note that the asymmetry across the contact and a neighboring magnetic layer may cause in-plane Rashba spin-orbit interactions. Within the ferromagnetic layer, the spin-splitting energy is expected to be much larger than the strength of the possible in-plane Rashba spin-orbit interaction, so the effect of the latter can be quenched. In the non-magnetic contact, the Rashba interaction can cause spin depolarization, which can play a deleterious role on the functionalities such as spin filtering and spin pumping that aim at harvesting the transmitted or reflected spin current. Nevertheless, concerning the functionality of the spin transfer torque, it is realized by the multiple scattering of electron between the two magnetic layers upon spin-unpolarized incidence, which is not affected by the Rashba interaction in the contacts.

## Summary

We have shown that atomically thin magnetic devices can feature universally a giant spin transfer torque, owning to the multiple reflections of electrons in the quantum coherent tunneling through the magnets. This finding introduces the classical paradigm of cavity quantum electrodynamics into magnetization manipulations, where the interaction of a flying object with a localized one is enhanced through multiple reflections inside the cavity. The roles of atoms and flying photon in cavity QED are now played by the magnets and the current carrying electron respectively, where van der Waals interfaces naturally realize the "cavity mirrors". Electron's functionality as an intermediate for angular momentum exchange between ferromagnetic layers is radically amplified by its multiple impacts on the magnets. The torque generated per electron incidence can become orders larger than $\hbar/2$ at the tunneling resonance. This regime has not been possible for conventional magnetic tunneling junction where the overall thickness exceeds the coherence length.

This mechanism of amplifying spin transfer torque is also distinguished from existing scheme exploiting the resonance tunneling through engineered barrier[31,32], where magnets are outside the multi-reflection path and hence torque generated per electron incidence is always upper bounded by $\hbar/2$. In fact, the torque amplification here is not fully relying on the tunneling resonance. Over a rather broad energy range around the tunneling resonance, we find the figure of merit – torque generated per electron tunneling – to have a sizable and universal value $(\hbar/2)\tan(\theta/2)$, determined by the angle $\theta$ between the magnetizations only, independent of model parameters.

**Acknowledgments:** The work is supported by the Research Grants Council of Hong Kong (Grants No. HKU17303518 and No. C7036-17W), and the University of Hong Kong (Seed Funding for Strategic Interdisciplinary Research).